\documentclass[twocolumn,aps,prd,superscriptaddress,nofootinbib]{revtex4-1}
\usepackage{graphicx}
\usepackage{color}
\graphicspath{{figures/}{fig/}}
\usepackage{amsmath}
\usepackage{amssymb}
\usepackage{bm}
\usepackage{slashed}
\usepackage{epsfig}
\usepackage{amsfonts}
\usepackage{epstopdf}
\usepackage{hyperref}
\usepackage{bbm}
\usepackage{textcomp}

\begin{document}
\title{Complete reduction of integrals in two-loop five-light-parton scattering amplitudes}

\author{Xin Guan}
\email{guanxin0507@pku.edu.cn}
\affiliation{School of Physics and State Key Laboratory of Nuclear Physics and
Technology, Peking University, Beijing 100871, China}

\author{Xiao Liu}
\email{xiao6@pku.edu.cn}
\affiliation{School of Physics and State Key Laboratory of Nuclear Physics and
Technology, Peking University, Beijing 100871, China}

\author{Yan-Qing Ma}
\email{yqma@pku.edu.cn}
\affiliation{School of Physics and State Key Laboratory of Nuclear Physics and
Technology, Peking University, Beijing 100871, China}
\affiliation{Center for High Energy Physics, Peking University, Beijing 100871, China}
\affiliation{Collaborative Innovation Center of Quantum Matter,
Beijing 100871, China}

\date{\today}

\begin{abstract}
We reduce all the most complicated Feynman integrals in two-loop five-light-parton scattering amplitudes to basic master integrals, while other integrals can be reduced even easier. Our results are expressed as systems of linear relations in the block-triangular form, very efficient for numerical calculations. Our results are crucial for complete next-to-next-to-leading order quantum chromodynamics calculations for three-jet, photon, and/or hadron production at hadron colliders. To determine the block-triangular relations, we develop an efficient and general method, which may provide a practical solution to the bottleneck problem of reducing multiloop multiscale integrals.

\end{abstract}

\maketitle
\allowdisplaybreaks

\section{Introduction}
 Owing to the good performance of the Large Hadron Collider (LHC), we have entered the era of precision high energy physics. Some of the most important observables are three light particles or jet production cross sections~\cite{Aaboud:2017lxm,Aaboud:2017fml,Sirunyan:2018adt}, which can both be used for testing the strong interaction at high energy and for determining the QCD coupling constant. From the theoretical viewpoint, predictions with compatible precision are needed, which requires perturbative QCD calculations up to next-to-next-to-leading order (NNLO). Although significant advances have been made in the past few years~\cite{Badger:2013gxa,Badger:2015lda,Badger:2017jhb,Abreu:2017hqn,Badger:2018enw,Abreu:2018jgq,Abreu:2018aqd, Boels:2018nrr,Abreu:2018zmy,Chicherin:2018yne,Chicherin:2019xeg,Abreu:2019rpt, Abreu:2019odu,Badger:2019djh,Hartanto:2019uvl,Chawdhry:2019bji,Gehrmann:2015bfy,Papadopoulos:2015jft,Gehrmann:2018yef, Chicherin:2018mue,Chicherin:2018old}, a complete NNLO result is still unavailable. One of the main concurrent obstacles is computation of two-loop amplitudes.

To evaluate a two-loop five-light-parton scattering amplitude, one usually first generates an integrand, reduces all of the Feynman integrals to linear combinations of relatively simpler master integrals (MIs), and finally calculates these MIs. Because integrands can be obtained either using the unitarity method~\cite{Badger:2013gxa,Badger:2015lda,Badger:2017jhb,Abreu:2017hqn,Badger:2018enw,Abreu:2018jgq} or using the conventional Feynman diagram method, and because MIs can be calculated analytically~\cite{Gehrmann:2015bfy,Papadopoulos:2015jft,Gehrmann:2018yef, Chicherin:2018mue,Chicherin:2018old}, the bottleneck is the reduction of Feynman integrals.  For example, the non-planar contribution of two-loop three-photon production at the LHC cannot be calculated, owing to the lack of such reduction for nonplanar integrals~\cite{Chawdhry:2019bji}.

Reduction is usually achieved by integration-by-parts (IBP) identities combined with Laporta's algorithm~\cite{Chetyrkin:1981qh,Laporta:2001dd,Smirnov:2008iw,Smirnov:2014hma,Smirnov:2019qkx, Maierhoefer:2017hyi, Maierhofer:2018gpa,Studerus:2009ye,vonManteuffel:2012np,Lee:2012cn,Peraro:2019svx}.
Although many interesting proposals have been made recently for improving the IBP reduction~\cite{ vonManteuffel:2014ixa,Peraro:2016wsq,Kosower:2018obg,Wang:2019mnn,Mastrolia:2018uzb, Frellesvig:2019kgj,Frellesvig:2019uqt,Klappert:2019emp,Gluza:2010ws,Schabinger:2011dz,Larsen:2015ped,Boehm:2018fpv,Bendle:2019csk,Chawdhry:2018awn}, the problem of reducing multiloop multiscale integrals has not been fully resolved yet.  The difficulty is twofold. On the one hand, owing to the number of scales, an explicit solution of the IBP system is usually too big to be used in numerical calculations; in addition, it is very difficult to obtain~\cite{Boehm:2018fpv,Bendle:2019csk,Chawdhry:2018awn,Borowka:2016ehy,Jones:2018hbb}. On the other hand, although solving the IBP system numerically in a single run is feasible, one usually needs to solve it many times, for either the phase space integration or fitting analytical expressions, which is very time- and resource-consuming. For example, to reconstruct the fully analytical two-loop five-gluon all-plus helicity amplitude~\cite{Badger:2019djh}, one needs to run the numerical computation of the IBP for nearly half a million times~\footnote{We thank Y. Zhang for pointing out this. Here and in the rest of the paper, if not specified, ``numerical" means rational numbers over a finite field of a big prime number.}. If one uses the same method to reconstruct analytical one-minus or maximal-helicity-violation amplitude, many more IBP calculation runs may be needed, which becomes prohibitive.

We note that a reduction can be obtained efficiently if a system of block-triangular relations is found, which has a small expression size and can be solved numerically very efficiently. Using our proposed series representation of Feynman integrals as input~\cite{Liu:2017jxz,Liu:2018dmc}, in Ref.~\cite{Liu:2018dmc} we described an algorithm that searched for block-triangular relations and yielded some preliminary results. Although our method developed in Ref.~\cite{Liu:2018dmc} is sufficiently good for reducing integrals with integrands having only denominators, the method is very time-consuming for physical problems that contain integrands with numerators.

In this paper, by further developing the method in Ref.~\cite{Liu:2018dmc}, we propose a two-step search strategy along with a reduction scheme that is suitable for physical problems. Based on this, we successfully find out block-triangular relations to reduce integrals in two-loop five-light-parton scattering amplitudes. As expected, the relations are only 148MB in size, and can be numerically solved hundreds of times faster than using other methods. Our work constitutes an important step towards the complete NNLO QCD calculation for three-jet, photon, or hadron production at the LHC. Because our method is efficient and general, it can be straightforwardly applied to any other process, thus providing a practical solution for the bottleneck problem of reducing Feynman integrals.

\section{Feynman integrals in two-loop five-light-parton scattering amplitudes}
\begin{figure}[htb]
	\begin{center}
		\includegraphics[width=0.8\linewidth]{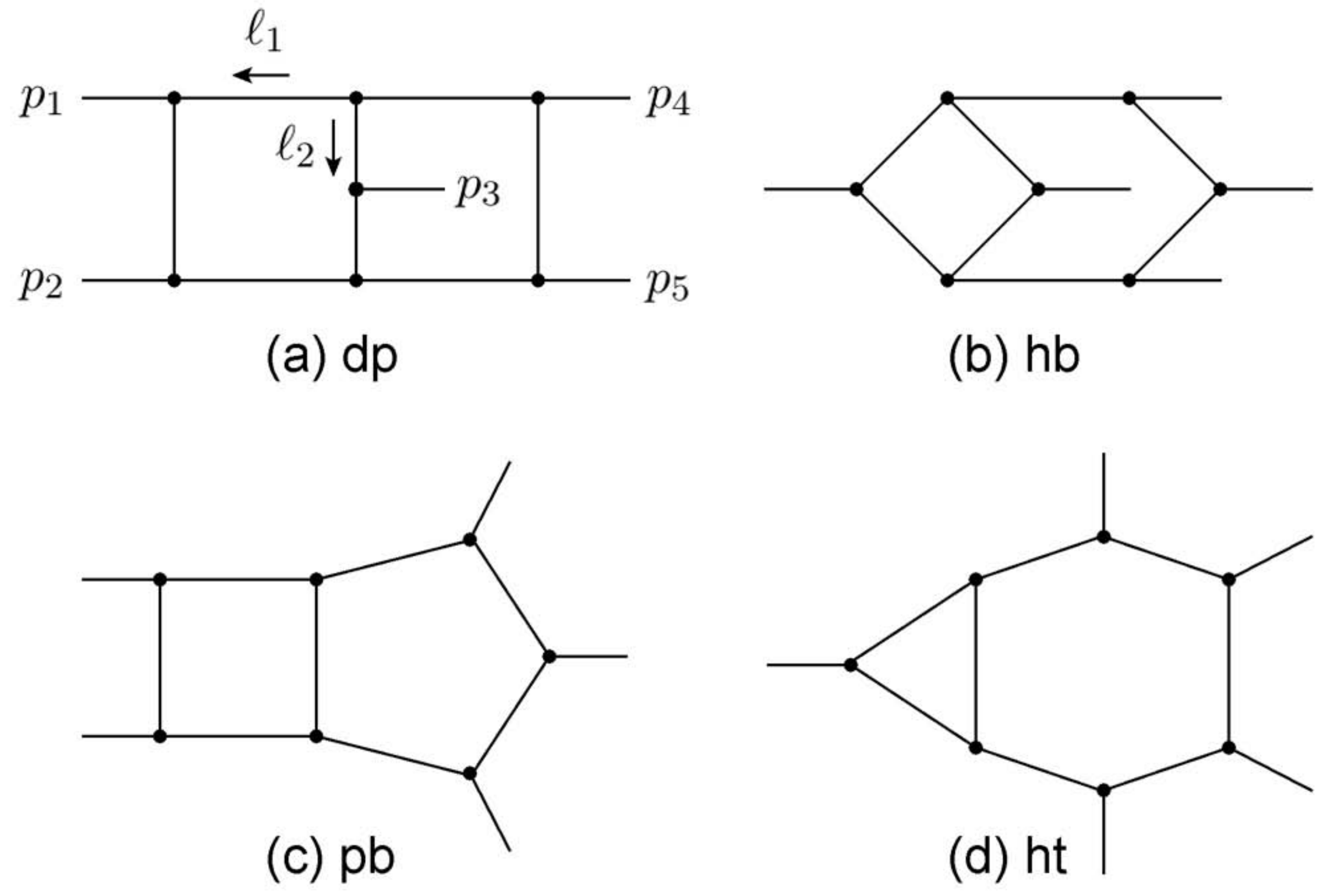}
		\caption{\label{fig:families}
			All 8-propagator families: (a) double-pentagon; (b) hexa-box; (c) penta-box; (d) hexa-triangle.}
	\end{center}
\end{figure}

To obtain the very much needed reduction of Feynman integrals in two-loop five-light-parton scattering amplitudes, we only need to consider integrals originated from the four topologies shown in Fig.~\ref{fig:families}. All the other Feynman integrals are one-loop-like, and can be dealt with much easier.

Let us consider the most complicated case, topology (a) in Fig.~\ref{fig:families}, as an example that will explain what kind of Feynman integrals do we need to reduce. There are five external momenta $p_1,\cdots,p_5$ flowing into the diagram, satisfying on-shell conditions $p_i^2=0$ ($i=1,\ldots,5$) and momentum conservation $\sum_{i=1}^5 p_i=0$. As a result, this problem contains five independent mass scales, which can be chosen as $\vec{s}=\{s_1,s_{2},s_{3},s_{4},s_{5}\}$ with $s_{i}\equiv2p_i\cdot p_{i+1}$ and $p_6\equiv p_1$.
With two loop momenta $\ell_1$ and $\ell_2$, a complete set of Lorentz scalars can be chosen as
\begin{align}\label{eq:Denominators}
	&D_1=\ell_1^2,D_2=(\ell_1+p_1)^2,\,D_3=(\ell_1+p_1+p_2)^2,\,\nonumber\\
	&D_4=\ell_2^2,\,D_5=(\ell_2+p_3)^2, D_6=(\ell_1+\ell_2+p_1+p_2+p_3)^2,\,\nonumber\\
	&D_7=(\ell_1+\ell_2-p_4)^2,\,D_8=(\ell_1+\ell_2)^2,\nonumber\\
	&D_9=(\ell_2+p_1)^2,D_{10}=(\ell_2+p_2)^2,D_{11}=(\ell_2+p_4)^2,
\end{align}
where the first eight are inverse propagators and the last three are introduced to make the set complete. Then the {\it family} of integrals defined by topology (a) can be expressed as
\begin{align}\label{eq:defintegral} I_{\vec{\nu}}(\epsilon,\vec{s}\,)=\int\frac{\text{d}^{4-2\epsilon}\ell_1\, \text{d}^{4-2\epsilon}\ell_2}{(\text{i}\pi^{2-\epsilon})^2}\,\frac{D_9^{-\nu_9}D_{10}^{-\nu_{10}}D_{11}^{-\nu_{11}}}{D_1^{\nu_1}...\,D_8^{\nu_8}},
\end{align}
where the indexes $\nu_{1}, \cdots, \nu_8$ are integers, $\nu_9$, $\nu_{10}$ and $\nu_{11}$ are nonpositive integers. Two integrals in this family are said to be in the same {\it sector} if the positions of their positive indexes are the same. The {\it degree} of an integral is defined by the opposite value of the summation of all its negative indexes. Finally, we call a degree-$m$ integral is $\frac{m}{n}$-{\it{type}} if it has $n$ positive indexes and all these positive indexes are $1$. For example, $I_{\{1,1,1,1,1,1,1,1,-4,0,-1\}}$ is a degree-5 integral in the top sector, and it is $\frac{5}{8}$-type.

For later convenience, we define operators ${\hat{m}^\pm}$ (for a non-negative integer $m$), which generate a set of integrals in the same sector or its subsectors when acting on an integral. For any integral $I_{\vec{\nu}}$, ${\hat{0}^\pm}I_{\vec{\nu}}=I_{\vec{\nu}}$, ${\widehat{m+1}^\pm}I_{\vec{\nu}}={\hat m^\pm \hat 1^\pm}I_{\vec{\nu}}$ , ${\hat 1^-}I_{\vec{\nu}}$ generates a set of integrals with one index decreased by 1, and ${\hat 1^+}I_{\vec{\nu}}$ generates a set of integrals with one nonzero index increased by 1. For example, we have
\begin{align}
\hat{1}^+&I_{\{1,1,1,1,1,1,1,1,-4,0,-1\}}=\{I_{\{2,1,1,1,1,1,1,1,-4,0,-1\}},\nonumber\\ &I_{\{1,2,1,1,1,1,1,1,-4,0,-1\}}, I_{\{1,1,2,1,1,1,1,1,-4,0,-1\}},\nonumber\\ &I_{\{1,1,1,2,1,1,1,1,-4,0,-1\}}, I_{\{1,1,1,1,2,1,1,1,-4,0,-1\}},\nonumber\\
& I_{\{1,1,1,1,1,2,1,1,-4,0,-1\}}, I_{\{1,1,1,1,1,1,2,1,-4,0,-1\}}, \nonumber\\ &I_{\{1,1,1,1,1,1,1,2,-4,0,-1\}}, I_{\{1,1,1,1,1,1,1,1,-3,0,-1\}},\nonumber\\
& I_{\{1,1,1,1,1,1,1,1,-4,0,0\}}\}\,,
\end{align}
and
\begin{align}
\hat{1}^-&I_{\{1,1,1,1,1,1,1,1,-4,0,-1\}}=\{I_{\{0,1,1,1,1,1,1,1,-4,0,-1\}},\nonumber\\
& I_{\{1,0,1,1,1,1,1,1,-4,0,-1\}}, I_{\{1,1,0,1,1,1,1,1,-4,0,-1\}}, \nonumber\\
&I_{\{1,1,1,0,1,1,1,1,-4,0,-1\}}, I_{\{1,1,1,1,0,1,1,1,-4,0,-1\}},\nonumber\\
& I_{\{1,1,1,1,1,0,1,1,-4,0,-1\}}, I_{\{1,1,1,1,1,1,0,1,-4,0,-1\}},  \nonumber\\
&I_{\{1,1,1,1,1,1,1,0,-4,0,-1\}}, I_{\{1,1,1,1,1,1,1,1,-5,0,-1\}},\nonumber\\
& I_{\{1,1,1,1,1,1,1,1,-4,-1,-1\}},I_{\{1,1,1,1,1,1,1,1,-4,0,-2\}}\}\,.
\end{align}
We also define operators ${\hat m}^\circleddash$,  which can generate a set of integrals as a union of integrals generated by $\{\hat m^-, \widehat{m-1}^-,\cdots,\hat 0^-\}$ when acting on an integral.

As is well-known, the {\it most complicated}\footnote{The definition of complexity is a consequence of a convention to order integrals. In our convention, integrals are thought to be more complicated if they have more propagators, integrals in the same sector are more complicated if they have higher total denominator powers or if they have higher degree, and so on.} integrals in the amplitudes are those with the highest number of propagators, i.e., $\nu_{i}=1~(i=1,\cdots,8)$, and the highest numerator degree, i.e., $-(\nu_9+\nu_{10}+\nu_{11})$. By studying the two-loop five-gluon scattering amplitude diagram by diagram, we find the highest numerator degree is 5 for all integrals. Therefore we define an integral set
\begin{align}\label{eq:Sa}
S_{(a)}={\hat 5}^\circleddash I_{\{1,1,1,1,1,1,1,1,0,0,0\}},
\end{align}
which contains 3914 nonzero integrals with all the most complicated integrals in five-gluon scattering amplitude being included. Because the five-gluon scattering amplitude is sufficiently general, all the most complicated integrals (if not all integrals) belonging to topology (a) appearing in five-light-parton scattering amplitudes are included in the set $S_{(a)}$. In fact, for two-loop five-gluon all-plus helicity amplitude, integrals in topology (a) form a subset of $S_{(a)}$ \cite{Badger:2015lda}. Therefore, for the purpose of reducing integrals in physical amplitudes,  the main job for topology (a) is to reduce integrals in set $S_{(a)}$.

For topologies (b), (c) and (d) in Fig.~\ref{fig:families}, we define sets of target integrals $S_{(b)}$, $S_{(c)}$ and $S_{(d)}$, similar to $S_{(a)}$.

\section{Search for block-triangular relations}
Before presenting our method for reducing two-loop five-light-parton integrals, let us first point out that for multiscale problems, expressing general integrals in terms of MIs explicitly is not preferred, even at the one-loop level. Instead, one usually sets up a system of block-triangular relations that can numerically relate all of the integrals to MIs (see \cite{Denner:2005nn} and references therein).

The advantage of a system of block-triangular relations over the explicit solution can be understood based on the integrals' singularities. If we express a complicated integral as a linear combination of simpler MIs, powers of Gram determinants will appear in the denominators of the coefficients of these MIs, which is  necessary because only thus the linear combination of MIs can generate correct singularities of the target integral. Then, the numerators of these coefficients will have high mass dimensions and thus will have very long expressions.  This difficulty can be nicely resolved using a system of block-triangular relations. Relations in each block can be very simple, but their solution can naturally generate Gram determinants in the denominator. Furthermore, correctly choosing the blocks may result in the solution involving only one Gram determinant.

Because reduction at multiloop level is much more complicated than for the one-loop case, the above discussion implies that constructing a system of block-triangular relations may be the best way to reduce multiloop multiscale integrals. Unlike one-loop case, where block-triangular systems can be achieved easily by analytically solving the IBP relations, block-triangular systems at multiloop level are in general difficult to obtain.

In Ref.~\cite{Liu:2018dmc}, based on our proposed series representation of Feynman integrals~\cite{Liu:2017jxz,Liu:2018dmc}  as input information, we constructed an algorithm that searched for block-triangular relations to reduce multiloop multiscale integrals. However, we found the method to be very time-consuming for physical problems, although  it was efficient for reducing integrals with integrands containing only denominators. To deal with physical problems such as two-loop five-light-parton integrals, we propose here a two-step search strategy.

In the first step, we set up a system of relations that can numerically express all target integrals in terms of MIs. The system is allowed to be somewhat inefficient in numerical calculations; thus, the system is not required to be block-triangular. This system can be obtained either by using our series representation of Feynman integrals~\cite{Liu:2018dmc}, or simply by using the well-known IBP system.

In the second step, we search for a system of block-triangular relations, which needs to be very efficient for numerical computations. The algorithm is the same as that proposed in Ref.~\cite{Liu:2018dmc} except that, instead of using our series representation of Feynman integrals, we use the numerical solution obtained in the first step as input information.

More details about the search strategy can be found in appendix.

\section{Reduction scheme and results}
To apply the above proposed search strategy on physical problems, we still need to introduce the reduction scheme, which amounts to choosing target integrals and other integrals that are allowed to appear in each block. In this paper, integrals in each block are defined by operator ${\hat m}^\circleddash$ acting on a proper integral. For example, to reduce the integrals in $S_{(a)}$, all of the integrals are allowed to appear in the first block, and the target integrals in this block are all the 21 most complicated integrals in the top sector with degree 5. The first block enables us to express all the 21 most complicated integrals in terms of simpler integrals. Then, in the second block, we choose the most complicated integrals among the rest of the integrals as target integrals, and use operator ${\hat m}^\circleddash$ acting on a proper integral to generate a set of integrals that covers all the target integrals. Then, the process is repeated. Eventually, any integral can be expressed in terms of simpler integrals.

\begin{center}
	\begin{table}[htb]
		\centering
		\begin{tabular}{|c|c|c|c|c|c|c|}
			\hline
			top.&$\#$int.&$\#$MIs&$t_{\text{search}}$ (h) & $t_{\text{solve}}$ (s)&size(MB)\\\hline
			(a) &$3914$  &$108$  &112  &0.17  &66    \\
			(b) &$3584$  &$73$   &31   & 0.090   &40 \\
			(c) &$3458$  &$61$   &56    & 0.075  &31 \\
			(d) &$2634$  &$28$   &8    & 0.035  &11 \\\hline
		\end{tabular}
		\caption{Main information of the obtained reduction relations. $t_{\text{search}}$ represents the CPU time required to search for these relations in the unit of CPU-core hours. $t_{\text{solve}}$ represents the time spent to solve these relations numerically using one CPU.}
		\label{tab:database}
	\end{table}
\end{center}

\begin{figure}[htb]
	\begin{center}
		\includegraphics[width=0.8\linewidth]{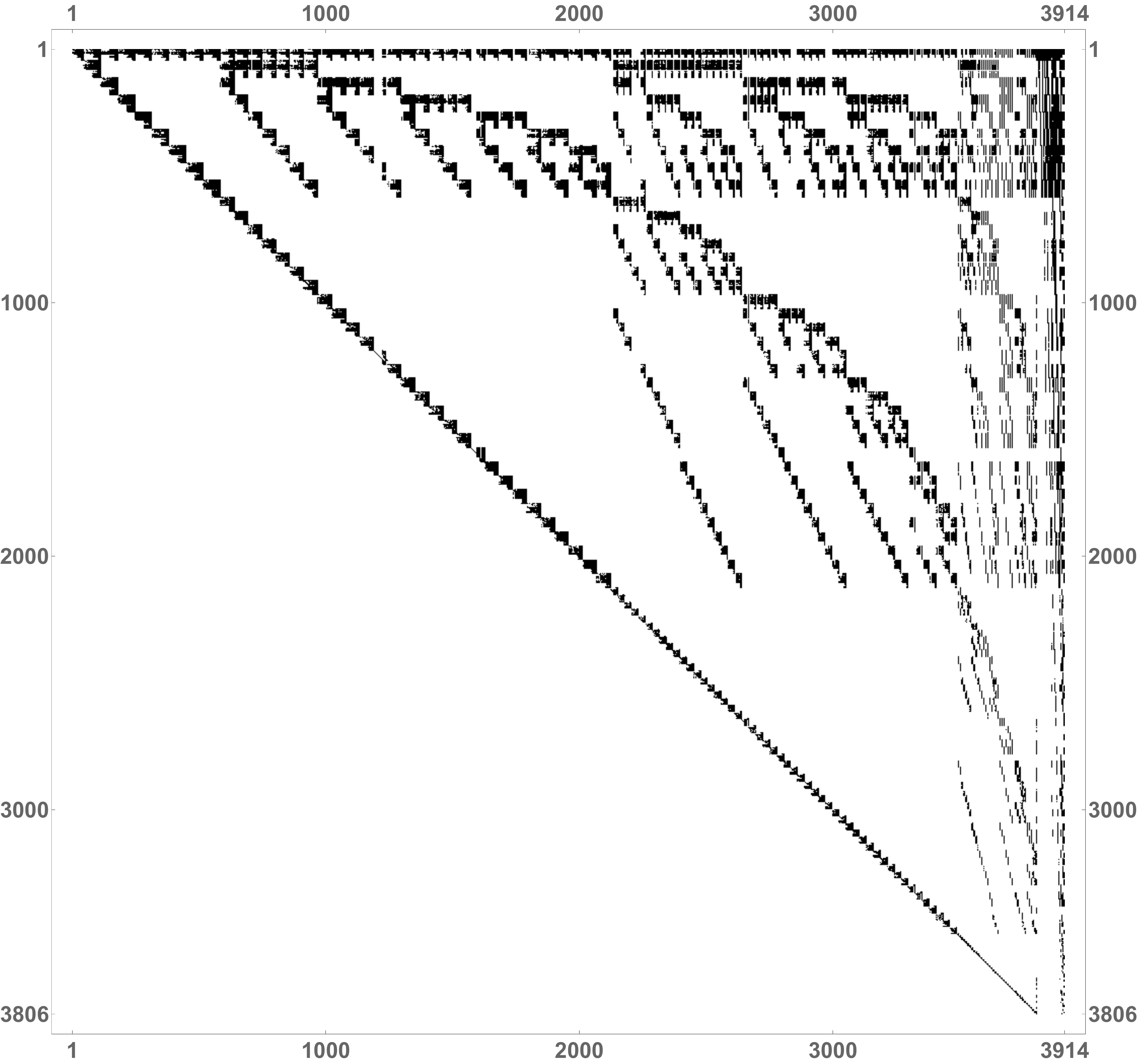}
		\caption{\label{fig:matrix}
			Matrix density plot for relations of topology (a). Each row represents a relation and each column represents an integral. Black points represent nonzero elements.}
	\end{center}
\end{figure}

Using the above method, we successfully determined systems of block-triangular relations for integrals in the four topologies in Fig.~\ref{fig:families}. The file sizes of all these relations are acceptable, $\sim$148 MB.  To obtain these results required $\sim$200 central processing unit (CPU) core hours to search for relations in the second step of the two-step search strategy, in addition to hundreds of CPU-core hours for generating input information by numerically solving the system obtained in the first step. Some basic information about these results is listed in Tab.~\ref{tab:database}.

For more intuitive understanding, we show a matrix density plot for the block-triangular system of topology (a) in Fig.~\ref{fig:matrix}. This system contains 3914 integrals and 108 MIs, which means we need 3806 linear relations to reduce all of the target integrals. In this plot, each line represents a relation, each column corresponds to an integral, and black points represent nonzero elements in the matrix. Integrals are ordered, from the most complicated one to the simplest one, with MIs at the end of each line. The matrix is exactly block-triangular, and the largest block contains only tens of relations.

Analytic expressions for all of these relations are available from the website in \cite{www:reduction}. Technical details of our reduction scheme can be found in appendix.

\section{validation and comparison with other methods}
Our final reduction relations have been verified numerically using an independent code  \verb'FIRE6'~\cite{Smirnov:2019qkx} for randomly chosen phase space points, and the results of both approaches were in a good agreement.

For each given numerical point $\epsilon$ and $\vec{s}$, solving our reduction relations of the four families cost 0.4 s using one CPU, as is shown in Tab.~\ref{tab:database}.
The time spent can be divided into two parts: assignment (substituting numerical $\epsilon$ and $\vec{s}$ into the system), which is proportional to the file size; and solving the system, which depends on both the number of relations and how these relations are coupled with each other. Because our systems are block-triangular, the time spent on the latter part is shorter. Therefore, the efficiency of numerical calculation of our reduction relations can be simply estimated by the file size.

Compared with explicit solutions, the file sizes of our reduction relations are much smaller. The file size for explicit solutions of eight-propagator integrals with degree up to 4 in topology (a), 26 integrals overall, is $\sim$2GB~\cite{Bendle:2019csk}; that for the explicit solutions of eight-propagator integrals with degree up to 4 in topology (b), 32 integrals overall, is $\sim$0.8GB~\cite{Boehm:2018fpv}; and that for the solutions of all integrals in topology (c) is in excess of 20GB for compressed format~\cite{Chawdhry:2018awn}. It can be expected that our relations should be hundreds of times smaller than the complete explicit solution in terms of the file size, which results in more than a 100-fold speedup of numerical calculations, even if there is no memory deficit for storing the huge expression of explicit solutions.

We note that the file size of trimmed IBP relations to reduce all of the integrals considered in this work is a few GB, which is also much larger than that of our reduction relations. The reason is that, although each IBP relation is simpler than ours, the IBP system involves hundreds of times more equations. Furthermore, the time spent on numerical IBP is dominated by the latter part because IBP relations are coupled in a complicated way.
As a result, numerical IBP should be much more inefficient than our method. Through our test, numerical IBP via \verb"FiniteFlow"~\cite{Peraro:2019svx} combined with \verb"LiteRed"~\cite{Lee:2012cn} costs about 2 minutes for each phase space point, which is slower than our method by more than a 100-fold.

The above comparison reveals the advantage of our method. Numerical evaluation of explicit solutions spends too much time on assignments; while numerical IBP spends too much time on solving linear equations. Our method performs better on both parts, and therefore it is much more efficient. Similar to numerical evaluation over the field of prime numbers, our reduction relations should also be much more efficient for numerical evaluations with floating numbers, which enables phase space integration to obtain physical cross sections.

\section{Summary and outlook}
In this paper, we achieved the reduction of a set of integrals which covers all of the most complicated integrals in two-loop five-light-parton scattering amplitudes. Our results are expressed as systems of linear relations in the block-triangular form, which are very efficient for numerical calculations. The remaining integrals involved in amplitudes can be easily reduced using the same method, on demand. Therefore, a complete reduction of integrals in two-loop five-light-parton scattering amplitudes, which challenges all other methods, is available now. Because MIs are already known~\cite{Gehrmann:2015bfy,Papadopoulos:2015jft,Gehrmann:2018yef, Chicherin:2018mue,Chicherin:2018old}, our results provide the complete calculation of two-loop five-light-parton scattering amplitudes, and thus complete NNLO calculation of three light particles or jet-production at the LHC on the horizon.

To obtain the block-triangular relations, we developed the method in Ref.~\cite{Liu:2018dmc} by proposing a two-step search strategy along with a reduction scheme. As our newly developed method is general and efficient, other more complicated problems, like two-loop integrals for $t\bar{t}+\text{jet}$, $t\bar{t}H$, or 4-jet hadron production, are also within reach. Our work opens the door for complete NNLO QCD calculations for production of three or more particles at the LHC.

In the current application of our method, most CPU time is allocated to solving the system obtained in the first step. Although the time spent is tolerable for the current problem, improvement may be needed for more complicated applications. There are different options. Using the method in \cite{Liu:2018dmc}, better integral sets can be explored. Another possible choice is to use trimmed IBP systems obtained by solving syzygy equations~\cite{Gluza:2010ws,Schabinger:2011dz,Larsen:2015ped,Boehm:2018fpv,Bendle:2019csk}. These possibilities will be addressed in future studies.

\begin{acknowledgments}
	We thank K.T. Chao, F. Feng, Q.J. Jin, Z. Li, X.H. Liu, H. Luo, C. Meng, J. Usovitsch and Y. Zhang for many useful communications and discussions.
	The work is supported in part by the National Natural Science Foundation of China (Grants No. 11875071, No. 11975029) and the High-performance
	Computing Platform of Peking University.
\end{acknowledgments}

\appendix

\section*{Appendix: Reduction method}

\subsection{Search strategy: step one}
We take integrals originated from topology (a) in the main text as an example for explaining the details of our technique.

We want to set up a set of relations, using which we can express all integrals in $S_{(a)}$ in terms of MIs for any given phase space point (rational numbers for both $\vec{s}$ and $\epsilon$), with coefficients calculated in the finite field of a 63-bit prime number. Although the IBP method~\cite{Chetyrkin:1981qh,Laporta:2001dd,Smirnov:2008iw,Smirnov:2014hma,Smirnov:2019qkx, Maierhoefer:2017hyi, Maierhofer:2018gpa,Studerus:2009ye,vonManteuffel:2012np,Lee:2012cn,Peraro:2019svx} can do this, we would like to explain in the following that our method proposed in ~\cite{Liu:2018dmc} may provide a better choice.

For each given integral $I_{\vec{\nu}}$, called a {\it seed}, there are 12 IBP relations among the integral set
\begin{align}\label{eq:GIBP}
G_{\vec{\nu}}^{\text{IBP}}=\{\hat{1}^+, \hat{1}^-\hat{1}^+\} I_{\vec{\nu}}.
\end{align}

In addition, there are 6 relations owing to the Lorentz invariance \cite{Gehrmann:1999as}, which can be interpreted as linear combinations of IBP relations from other seeds \cite{Lee:2008tj}.

The above IBP relations can also be found out easily using the method proposed in ~\cite{Liu:2018dmc}. To this end, we introduce a parameter $\eta$ for all integrals in $G_{\vec{\nu}}^{\text{IBP}}$, and then search relations among them using input information from the series representation \cite{Liu:2017jxz,Liu:2018dmc}. Up to $d_{\text{max}}=1$, where $d_{\text{max}}$ is a half of the maximal value of mass dimension for the coefficients of relations, we can find at least 12 relations; while up to $d_{\text{max}}=2$ we find at least 12+6 relations. Because these relations are analytical in $\eta$, we can take $\eta\to0$ directly and recover the aforementioned $12+6$ IBP relations.

The advantage of our method in ~\cite{Liu:2018dmc} is that it allows to search relations among any set of integrals. As the simplest generalization of $G_{\vec{\nu}}^{\text{IBP}}$, we can define an integral set
\begin{align}\label{eq:GIBPt}
G_{\vec{\nu}}=\{\hat{1}^+, \hat{1}^-\hat{1}^+,\hat{1}^-\} I_{\vec{\nu}},
\end{align}
and search relations among them. Up to $d_{\text{max}}=2$, there are typically 2 more relations besides 12+6 IBP relations for each seed. With more relations in hand, it is possible to select better relations to achieve a more efficient reduction. For example, our relations from all $\frac{4}{8}$-type seeds can already reduce 15 out of all $\frac{5}{8}$-type integrals to integrals with lower degree (these relations are available at \cite{www:reduction}). IBP relations from these seeds cannot achieve this because $\frac{5}{8}$-type integrals do not show up.

One can certainly explore other integral sets for each seed, to further improve the reduction efficiency. We did not do that because efficiency of either the IBP set~\eqref{eq:GIBP} or the generalized set~\eqref{eq:GIBPt} is sufficient for us to deal with the problem in this work.

With integral sets in hand, we generate a system of linear equations from all seeds belonging to $\frac{m}{n}$-type with $3\leq n \leq 8$ and $0\leq m \leq 5$, and use the package \verb"FiniteFlow"~\cite{Peraro:2019svx} to trim the system by removing redundant relations and solving the trimmed system numerically, which expresses all integrals in $S_{(a)}$ as linear combinations of 108 MIs (after exploring symmetries among MIs using \verb"LiteRed").

\subsection{Search strategy: step two}
In this step, we search linear relations to reduce the given target integrals in $G_1\subseteq S_{(a)}$ to simpler integrals in $G_2\subseteq S_{(a)}$ (the reducibility can be tested numerically easily). Combining the reduction scheme that will be described in the next section, a block-triangular system can be finally obtained.

We first describe how to search linear relations among the integral set $G:=\{I_1,\ldots,I_N\}\subseteq S_{(a)}$ of the form
\begin{align}\label{eq:eqform}
\sum_{i=1}^N Q_i(\epsilon,\vec{s}\,) I_i(\epsilon,\vec{s}\,)=0\,,
\end{align}
where $Q_i(\epsilon,\vec{s}\,)$ can be decomposed as
\begin{align}\label{eq:ExpandQ}
Q_i(\epsilon,\vec{s}\,) =\sum_{\kappa=0}^{\epsilon_{\text{max}}} \sum_{\vec{\lambda}\in\Omega_{d_i}} \tilde{Q}_i^{ \kappa\lambda_1\ldots \lambda_5} \,\epsilon^{\kappa}s_1^{\lambda_1}\cdots s_5^{\lambda_5},
\end{align}
where $\epsilon_\text{max}$ is the maximal power of $\epsilon$ allowed to appear in the relation, $\Omega_{d_i}=\{\vec{\lambda}\in\mathbb{N}^5|\,\lambda_1+\cdots+\lambda_5=d_i\}$, $d_i$ is half of the mass dimension of $Q_i$ which can be fixed by $d_{\text{max}}\equiv{\text{max}}\{d_1,\cdots,d_N\}$,  and $\tilde{Q}_i^{ \kappa\lambda_1\ldots \lambda_5}$ are unknown rational numbers to be determined. It is crucial to point out that, for given $\epsilon_{\text{max}}$ and $d_{\text{max}}$, the number of unknowns is finite. Therefore, it can be determined by a finite number of constraints. As will be explained in the following, these unknowns can be determined by the result obtained in the first step.

Based on the system of equations in the first step, for a given numerical point $\epsilon$ and $\vec{s}$ every integral in $G$ can be represented as an 108-dimensional vector, with elements being the projection onto MIs,
\begin{align}\label{eq:numsol}
I_i=\{C_{i,1}, \ldots, C_{i,108}\}\,,\quad i=1,\ldots,N\,.
\end{align}

By inserting these numerical vectors into Eq.~\eqref{eq:eqform}, we obtain a vector equation, which results in at most 108 independent constraints over the unknowns.  By repeating the above procedure many times (at most several thousand in this work), sufficient number of constraints can be obtained, for determining all of the unknowns. As the above values are actually calculated in the finite field of a given prime number, we still need to repeat the procedure for several different prime numbers (at most 15 in this work) and use the Chinese remainder theorem to reconstruct the real results of the unknowns. Finally, linear relations with given $d_\text{max}$ and $\epsilon_\text{max}$ are obtained.

To reduce $G_1$ to $G_2$, we just set $G:=G_1\cup G_2$ and search relations among $G$ with different values of $d_\text{max}$ and $\epsilon_{\text{max}}$. For the purpose of the current work, we find it is sufficient to fix $\epsilon_{\text{max}}=3$. To find out simple relations, we follow the algorithm proposed in ~\cite{Liu:2018dmc} by starting the search procedure with $d_\text{max}=0$ and increasing $d_\text{max}$ by 1 each time, until enough relations are obtained to reduce $G_1$ to $G_2$.

\subsection{Reduction scheme}
Reduction scheme determines which integrals should be involved in each block. We generate the integrals through previously defined operator ${\hat m}^\circleddash$ acting on properly chosen integrals.

For example, in the first block for topology (a), we need to reduce the most complicated $\frac{5}{8}$-type integrals. To this end, we set $G:=S_{(a)}={\hat 5}^\circleddash I_{\{1,1,1,1,1,1,1,1,0,0,0\}}$ with $G_1$ chosen as all 21 $\frac{5}{8}$-type integrals. We indeed find out 21 independent relations, which can reduce all $\frac{5}{8}$-type integrals to simpler integrals. The most complicated relation corresponds to $d_\text{max}=7$, which means that the coefficients of $\frac{5}{8}$-type integrals are degree-2 polynomials in $\vec{s}$. We then reduce $\frac{4}{8}$-type integrals, which can be realized by setting $G={\hat 4}^\circleddash I_{\{1,1,1,1,1,1,1,1,0,0,0\}}$ with $G_1$ chosen as all 15 $\frac{4}{8}$-type integrals. To reduce the rest of the top-sector integrals, we set $G={\hat 3}^\circleddash I_{\{1,1,1,1,1,1,1,1,0,0,0\}}$ with $G_1$ chosen as 11 top-sector integrals that are not MIs.

After reducing top-sector integrals, we still need to reduce integrals in subsectors. For example, for the seven-propagator sector $I_{\{1,1,1,1,1,1,1,0,0,0,0\}}$, whose most complicated integrals in $S_{(a)}$ are of $\frac{4}{7}$-type, we set $G={\hat 4}^\circleddash I_{\{1,1,1,1,1,1,1,0,0,0,0\}}$ with $G_1$ chosen as all 35 $\frac{4}{7}$-type integrals in this sector.

Based on the above scheme, we obtain 3801 reduction relations. By introducing additional 5 symmetry relations among MIs, we have 3806 relations in total that can express 3914 integrals in $S_{(a)}$ as linear combinations of 108 MIs.

We note that there is a way to further reduce the block size that has not been applied in this work. For example, by setting $G={\hat 3}^\circleddash I_{\{0,1,1,1,1,1,1,1,-1,0,0\}}$, we can generate smaller-size blocks to reduce a part of $\frac{4}{7}$-type integrals.

\providecommand{\href}[2]{#2}\begingroup\raggedright\endgroup


\begin{thebibliography}{10}

\bibitem{Aaboud:2017lxm}
{\bfseries ATLAS} , M.~Aaboud {\em et al.}, {\it {Measurement of the production
  cross section of three isolated photons in $pp$ collisions at $\sqrt{s}$ = 8
  TeV using the ATLAS detector}},
\href{http://dx.doi.org/10.1016/j.physletb.2018.03.057}{{\em Phys. Lett.}
  {\bfseries B781} (2018) 55--76}
  [\href{http://arxiv.org/abs/1712.07291}{{\ttfamily arXiv:1712.07291}}]
  [\href{http://inspirehep.net/search?p=find+Aaboud:2017lxm}{{\ttfamily
  InSPIRE}}].

\bibitem{Aaboud:2017fml}
{\bfseries ATLAS} , M.~Aaboud {\em et al.}, {\it {Determination of the strong
  coupling constant $\alpha _\mathrm {s}$ from transverse energy-energy
  correlations in multijet events at $\sqrt{s} = 8~\hbox {TeV}$ using the ATLAS
  detector}},
\href{http://dx.doi.org/10.1140/epjc/s10052-017-5442-0}{{\em Eur. Phys. J.}
  {\bfseries C77} (2017) 872}
  [\href{http://arxiv.org/abs/1707.02562}{{\ttfamily arXiv:1707.02562}}]
  [\href{http://inspirehep.net/search?p=find+Aaboud:2017fml}{{\ttfamily
  InSPIRE}}].

\bibitem{Sirunyan:2018adt}
{\bfseries CMS} , A.~M. Sirunyan {\em et al.}, {\it {Event shape variables
  measured using multijet final states in proton-proton collisions at $
  \sqrt{s}=13 $ TeV}},
\href{http://dx.doi.org/10.1007/JHEP12(2018)117}{{\em JHEP} {\bfseries 12}
  (2018) 117} [\href{http://arxiv.org/abs/1811.00588}{{\ttfamily
  arXiv:1811.00588}}]
  [\href{http://inspirehep.net/search?p=find+Sirunyan:2018adt}{{\ttfamily
  InSPIRE}}].

\bibitem{Badger:2013gxa}
S.~Badger, H.~Frellesvig, and Y.~Zhang, {\it {A Two-Loop Five-Gluon Helicity
  Amplitude in QCD}},
\href{http://dx.doi.org/10.1007/JHEP12(2013)045}{{\em JHEP} {\bfseries 12}
  (2013) 045} [\href{http://arxiv.org/abs/1310.1051}{{\ttfamily
  arXiv:1310.1051}}]
  [\href{http://inspirehep.net/search?p=find+Badger:2013gxa}{{\ttfamily
  InSPIRE}}].

\bibitem{Badger:2015lda}
S.~Badger, G.~Mogull, A.~Ochirov, and D.~O'Connell, {\it {A Complete Two-Loop,
  Five-Gluon Helicity Amplitude in Yang-Mills Theory}},
\href{http://dx.doi.org/10.1007/JHEP10(2015)064}{{\em JHEP} {\bfseries 10}
  (2015) 064} [\href{http://arxiv.org/abs/1507.08797}{{\ttfamily
  arXiv:1507.08797}}]
  [\href{http://inspirehep.net/search?p=find+Badger:2015lda}{{\ttfamily
  InSPIRE}}].

\bibitem{Badger:2017jhb}
S.~Badger, C.~Br$\o$nnum-Hansen, H.~B. Hartanto, and T.~Peraro, {\it {First
  look at two-loop five-gluon scattering in QCD}},
\href{http://dx.doi.org/10.1103/PhysRevLett.120.092001}{{\em Phys. Rev. Lett.}
  {\bfseries 120} (2018) 092001}
  [\href{http://arxiv.org/abs/1712.02229}{{\ttfamily arXiv:1712.02229}}]
  [\href{http://inspirehep.net/search?p=find+Badger:2017jhb}{{\ttfamily
  InSPIRE}}].

\bibitem{Abreu:2017hqn}
S.~Abreu, F.~Febres~Cordero, H.~Ita, B.~Page, and M.~Zeng, {\it {Planar
  Two-Loop Five-Gluon Amplitudes from Numerical Unitarity}},
\href{http://dx.doi.org/10.1103/PhysRevD.97.116014}{{\em Phys. Rev.} {\bfseries
  D97} (2018) 116014} [\href{http://arxiv.org/abs/1712.03946}{{\ttfamily
  arXiv:1712.03946}}]
  [\href{http://inspirehep.net/search?p=find+Abreu:2017hqn}{{\ttfamily
  InSPIRE}}].

\bibitem{Badger:2018enw}
S.~Badger, C.~Br$\o$nnum-Hansen, H.~B. Hartanto, and T.~Peraro, {\it {Analytic
  helicity amplitudes for two-loop five-gluon scattering: the single-minus
  case}},
\href{http://dx.doi.org/10.1007/JHEP01(2019)186}{{\em JHEP} {\bfseries 01}
  (2019) 186} [\href{http://arxiv.org/abs/1811.11699}{{\ttfamily
  arXiv:1811.11699}}]
  [\href{http://inspirehep.net/search?p=find+Badger:2018enw}{{\ttfamily
  InSPIRE}}].

\bibitem{Abreu:2018jgq}
S.~Abreu, F.~Febres~Cordero, H.~Ita, B.~Page, and V.~Sotnikov, {\it {Planar
  Two-Loop Five-Parton Amplitudes from Numerical Unitarity}},
\href{http://dx.doi.org/10.1007/JHEP11(2018)116}{{\em JHEP} {\bfseries 11}
  (2018) 116} [\href{http://arxiv.org/abs/1809.09067}{{\ttfamily
  arXiv:1809.09067}}]
  [\href{http://inspirehep.net/search?p=find+Abreu:2018jgq}{{\ttfamily
  InSPIRE}}].

\bibitem{Abreu:2018aqd}
S.~Abreu, L.~J. Dixon, E.~Herrmann, B.~Page, and M.~Zeng, {\it {The two-loop
  five-point amplitude in $\mathcal{N} =4$ super-Yang-Mills theory}},
\href{http://dx.doi.org/10.1103/PhysRevLett.122.121603}{{\em Phys. Rev. Lett.}
  {\bfseries 122} (2019) 121603}
  [\href{http://arxiv.org/abs/1812.08941}{{\ttfamily arXiv:1812.08941}}]
  [\href{http://inspirehep.net/search?p=find+Abreu:2018aqd}{{\ttfamily
  InSPIRE}}].

\bibitem{Boels:2018nrr}
R.~H. Boels, Q.~Jin, and H.~Luo,
{\it {Efficient integrand reduction for particles with spin}},
  [\href{http://arxiv.org/abs/1802.06761}{{\ttfamily arXiv:1802.06761}}]
  [\href{http://inspirehep.net/search?p=find+Boels:2018nrr}{{\ttfamily
  InSPIRE}}].

\bibitem{Abreu:2018zmy}
S.~Abreu, J.~Dormans, F.~Febres~Cordero, H.~Ita, and B.~Page, {\it {Analytic
  Form of Planar Two-Loop Five-Gluon Scattering Amplitudes in QCD}},
\href{http://dx.doi.org/10.1103/PhysRevLett.122.082002}{{\em Phys. Rev. Lett.}
  {\bfseries 122} (2019) 082002}
  [\href{http://arxiv.org/abs/1812.04586}{{\ttfamily arXiv:1812.04586}}]
  [\href{http://inspirehep.net/search?p=find+Abreu:2018zmy}{{\ttfamily
  InSPIRE}}].

\bibitem{Chicherin:2018yne}
D.~Chicherin, T.~Gehrmann, J.~M. Henn, P.~Wasser, Y.~Zhang, and S.~Zoia, {\it
  {Analytic result for a two-loop five-particle amplitude}},
\href{http://dx.doi.org/10.1103/PhysRevLett.122.121602}{{\em Phys. Rev. Lett.}
  {\bfseries 122} (2019) 121602}
  [\href{http://arxiv.org/abs/1812.11057}{{\ttfamily arXiv:1812.11057}}]
  [\href{http://inspirehep.net/search?p=find+Chicherin:2018yne}{{\ttfamily
  InSPIRE}}].

\bibitem{Chicherin:2019xeg}
D.~Chicherin, T.~Gehrmann, J.~M. Henn, P.~Wasser, Y.~Zhang, and S.~Zoia, {\it
  {The two-loop five-particle amplitude in $ \mathcal{N} $ = 8 supergravity}},
\href{http://dx.doi.org/10.1007/JHEP03(2019)115}{{\em JHEP} {\bfseries 03}
  (2019) 115} [\href{http://arxiv.org/abs/1901.05932}{{\ttfamily
  arXiv:1901.05932}}]
  [\href{http://inspirehep.net/search?p=find+Chicherin:2019xeg}{{\ttfamily
  InSPIRE}}].

\bibitem{Abreu:2019rpt}
S.~Abreu, L.~J. Dixon, E.~Herrmann, B.~Page, and M.~Zeng, {\it {The two-loop
  five-point amplitude in $ \mathcal{N} $ = 8 supergravity}},
\href{http://dx.doi.org/10.1007/JHEP03(2019)123}{{\em JHEP} {\bfseries 03}
  (2019) 123} [\href{http://arxiv.org/abs/1901.08563}{{\ttfamily
  arXiv:1901.08563}}]
  [\href{http://inspirehep.net/search?p=find+Abreu:2019rpt}{{\ttfamily
  InSPIRE}}].

\bibitem{Abreu:2019odu}
S.~Abreu, J.~Dormans, F.~Febres~Cordero, H.~Ita, B.~Page, and V.~Sotnikov, {\it
  {Analytic Form of the Planar Two-Loop Five-Parton Scattering Amplitudes in
  QCD}},
\href{http://dx.doi.org/10.1007/JHEP05(2019)084}{{\em JHEP} {\bfseries 05}
  (2019) 084} [\href{http://arxiv.org/abs/1904.00945}{{\ttfamily
  arXiv:1904.00945}}]
  [\href{http://inspirehep.net/search?p=find+Abreu:2019odu}{{\ttfamily
  InSPIRE}}].

\bibitem{Badger:2019djh}
S.~Badger, D.~Chicherin, T.~Gehrmann, G.~Heinrich, J.~M. Henn, T.~Peraro,
  P.~Wasser, Y.~Zhang, and S.~Zoia, {\it {Analytic form of the full two-loop
  five-gluon all-plus helicity amplitude}},
\href{http://dx.doi.org/10.1103/PhysRevLett.123.071601}{{\em Phys. Rev. Lett.}
  {\bfseries 123} (2019) 071601}
  [\href{http://arxiv.org/abs/1905.03733}{{\ttfamily arXiv:1905.03733}}]
  [\href{http://inspirehep.net/search?p=find+Badger:2019djh}{{\ttfamily
  InSPIRE}}].

\bibitem{Hartanto:2019uvl}
H.~B. Hartanto, S.~Badger, C.~Br$\o$nnum-Hansen, and T.~Peraro, {\it {A
  numerical evaluation of planar two-loop helicity amplitudes for a W-boson
  plus four partons}},
\href{http://dx.doi.org/10.1007/JHEP09(2019)119}{{\em JHEP} {\bfseries 09}
  (2019) 119} [\href{http://arxiv.org/abs/1906.11862}{{\ttfamily
  arXiv:1906.11862}}]
  [\href{http://inspirehep.net/search?p=find+Hartanto:2019uvl}{{\ttfamily
  InSPIRE}}].

\bibitem{Chawdhry:2019bji}
H.~A. Chawdhry, M.~L. Czakon, A.~Mitov, and R.~Poncelet,
{\it {NNLO QCD corrections to three-photon production at the LHC}},
\href{http://dx.doi.org/10.1007/JHEP02(2020)057} {{\em JHEP}
  {\bfseries 02} (2020) 057}
  [\href{http://arxiv.org/abs/1911.00479}{{\ttfamily arXiv:1911.00479}}]
  [\href{http://inspirehep.net/search?p=find+Chawdhry:2019bji}{{\ttfamily
  InSPIRE}}].

\bibitem{Gehrmann:2015bfy}
T.~Gehrmann, J.~M. Henn, and N.~A. Lo~Presti, {\it {Analytic form of the
  two-loop planar five-gluon all-plus-helicity amplitude in QCD}},
  \href{http://dx.doi.org/10.1103/PhysRevLett.116.189903,
  10.1103/PhysRevLett.116.062001}{{\em Phys. Rev. Lett.} {\bfseries 116} (2016)
  062001} [\href{http://arxiv.org/abs/1511.05409}{{\ttfamily
  arXiv:1511.05409}}]
  [\href{http://inspirehep.net/search?p=find+Gehrmann:2015bfy}{{\ttfamily
  InSPIRE}}].
[Erratum: Phys. Rev. Lett.116,no.18,189903(2016)].

\bibitem{Papadopoulos:2015jft}
C.~G. Papadopoulos, D.~Tommasini, and C.~Wever, {\it {The Pentabox Master
  Integrals with the Simplified Differential Equations approach}},
\href{http://dx.doi.org/10.1007/JHEP04(2016)078}{{\em JHEP} {\bfseries 04}
  (2016) 078} [\href{http://arxiv.org/abs/1511.09404}{{\ttfamily
  arXiv:1511.09404}}]
  [\href{http://inspirehep.net/search?p=find+Papadopoulos:2015jft}{{\ttfamily
  InSPIRE}}].

\bibitem{Gehrmann:2018yef}
T.~Gehrmann, J.~M. Henn, and N.~A. Lo~Presti, {\it {Pentagon functions for
  massless planar scattering amplitudes}},
\href{http://dx.doi.org/10.1007/JHEP10(2018)103}{{\em JHEP} {\bfseries 10}
  (2018) 103} [\href{http://arxiv.org/abs/1807.09812}{{\ttfamily
  arXiv:1807.09812}}]
  [\href{http://inspirehep.net/search?p=find+Gehrmann:2018yef}{{\ttfamily
  InSPIRE}}].

\bibitem{Chicherin:2018mue}
D.~Chicherin, T.~Gehrmann, J.~M. Henn, N.~A. Lo~Presti, V.~Mitev, and
  P.~Wasser, {\it {Analytic result for the nonplanar hexa-box integrals}},
\href{http://dx.doi.org/10.1007/JHEP03(2019)042}{{\em JHEP} {\bfseries 03}
  (2019) 042} [\href{http://arxiv.org/abs/1809.06240}{{\ttfamily
  arXiv:1809.06240}}]
  [\href{http://inspirehep.net/search?p=find+Chicherin:2018mue}{{\ttfamily
  InSPIRE}}].

\bibitem{Chicherin:2018old}
D.~Chicherin, T.~Gehrmann, J.~M. Henn, P.~Wasser, Y.~Zhang, and S.~Zoia, {\it
  {All Master Integrals for Three-Jet Production at Next-to-Next-to-Leading
  Order}},
\href{http://dx.doi.org/10.1103/PhysRevLett.123.041603}{{\em Phys. Rev. Lett.}
  {\bfseries 123} (2019) 041603}
  [\href{http://arxiv.org/abs/1812.11160}{{\ttfamily arXiv:1812.11160}}]
  [\href{http://inspirehep.net/search?p=find+Chicherin:2018old}{{\ttfamily
  InSPIRE}}].

\bibitem{Chetyrkin:1981qh}
K.~G. Chetyrkin and F.~V. Tkachov, {\it {Integration by Parts: The Algorithm to
  Calculate beta Functions in 4 Loops}},
\href{http://dx.doi.org/10.1016/0550-3213(81)90199-1}{{\em Nucl. Phys.}
  {\bfseries B192} (1981) 159--204}
  [\href{http://inspirehep.net/search?p=find+Chetyrkin:1981qh}{{\ttfamily
  InSPIRE}}].

\bibitem{Laporta:2001dd}
S.~Laporta, {\it {High precision calculation of multiloop Feynman integrals by
  difference equations}},
\href{http://dx.doi.org/10.1016/S0217-751X(00)00215-7,
  10.1142/S0217751X00002157}{{\em Int. J. Mod. Phys.} {\bfseries A15} (2000)
  5087--5159} [\href{http://arxiv.org/abs/hep-ph/0102033}{{\ttfamily
  hep-ph/0102033}}]
  [\href{http://inspirehep.net/search?p=find+Laporta:2001dd}{{\ttfamily
  InSPIRE}}].

\bibitem{Smirnov:2008iw}
A.~V. Smirnov, {\it {Algorithm FIRE -- Feynman Integral REduction}},
\href{http://dx.doi.org/10.1088/1126-6708/2008/10/107}{{\em JHEP} {\bfseries
  10} (2008) 107} [\href{http://arxiv.org/abs/0807.3243}{{\ttfamily
  arXiv:0807.3243}}]
  [\href{http://inspirehep.net/search?p=find+Smirnov:2008iw}{{\ttfamily
  InSPIRE}}].

\bibitem{Smirnov:2014hma}
A.~V. Smirnov, {\it {FIRE5: a C++ implementation of Feynman Integral
  REduction}},
\href{http://dx.doi.org/10.1016/j.cpc.2014.11.024}{{\em Comput. Phys. Commun.}
  {\bfseries 189} (2015) 182--191}
  [\href{http://arxiv.org/abs/1408.2372}{{\ttfamily arXiv:1408.2372}}]
  [\href{http://inspirehep.net/search?p=find+Smirnov:2014hma}{{\ttfamily
  InSPIRE}}].

\bibitem{Smirnov:2019qkx}
A.~V. Smirnov and F.~S. Chuharev,
{\it {FIRE6: Feynman Integral REduction with Modular Arithmetic}},
\href{https://doi.org/10.1016/j.cpc.2019.106877} {{\em Comput. Phys. Commun.}
  {\bfseries 247} (2020) 106877}
  [\href{http://arxiv.org/abs/1901.07808}{{\ttfamily arXiv:1901.07808}}]
  [\href{http://inspirehep.net/search?p=find+Smirnov:2019qkx}{{\ttfamily
  InSPIRE}}].

\bibitem{Maierhoefer:2017hyi}
P.~Maierhoefer, J.~Usovitsch, and P.~Uwer,
{\it {Kira - A Feynman Integral Reduction Program}},
\href{http://dx.doi.org/10.1016/j.cpc.2018.04.012} { {\em Comput. Phys. Commun.}
  {\bfseries 230} (2018) 99--112}
  [\href{http://arxiv.org/abs/1705.05610}{{\ttfamily arXiv:1705.05610}}]
  [\href{http://inspirehep.net/search?p=find+Maierhoefer:2017hyi}{{\ttfamily
  InSPIRE}}].

\bibitem{Maierhofer:2018gpa}
P.~Maierh\;ofer and J.~Usovitsch,
{\it {Kira 1.2 Release Notes}},
  [\href{http://arxiv.org/abs/1812.01491}{{\ttfamily arXiv:1812.01491}}]
  [\href{http://inspirehep.net/search?p=find+Maierhofer:2018gpa}{{\ttfamily
  InSPIRE}}].

\bibitem{Studerus:2009ye}
C.~Studerus, {\it {Reduze-Feynman Integral Reduction in C++}},
\href{http://dx.doi.org/10.1016/j.cpc.2010.03.012}{{\em Comput. Phys. Commun.}
  {\bfseries 181} (2010) 1293--1300}
  [\href{http://arxiv.org/abs/0912.2546}{{\ttfamily arXiv:0912.2546}}]
  [\href{http://inspirehep.net/search?p=find+Studerus:2009ye}{{\ttfamily
  InSPIRE}}].

\bibitem{vonManteuffel:2012np}
A.~von Manteuffel and C.~Studerus,
{\it {Reduze 2 - Distributed Feynman Integral Reduction}},
  [\href{http://arxiv.org/abs/1201.4330}{{\ttfamily arXiv:1201.4330}}]
  [\href{http://inspirehep.net/search?p=find+vonManteuffel:2012np}{{\ttfamily
  InSPIRE}}].

\bibitem{Lee:2012cn}
R.~N. Lee,
{\it {Presenting LiteRed: a tool for the Loop InTEgrals REDuction}},
  [\href{http://arxiv.org/abs/1212.2685}{{\ttfamily arXiv:1212.2685}}]
  [\href{http://inspirehep.net/search?p=find+Lee:2012cn}{{\ttfamily InSPIRE}}].

\bibitem{Peraro:2019svx}
T.~Peraro, {\it {FiniteFlow: multivariate functional reconstruction using
  finite fields and dataflow graphs}},
\href{http://dx.doi.org/10.1007/JHEP07(2019)031}{{\em JHEP} {\bfseries 07}
  (2019) 031} [\href{http://arxiv.org/abs/1905.08019}{{\ttfamily
  arXiv:1905.08019}}]
  [\href{http://inspirehep.net/search?p=find+Peraro:2019svx}{{\ttfamily
  InSPIRE}}].

\bibitem{vonManteuffel:2014ixa}
A.~von Manteuffel and R.~M. Schabinger, {\it {A novel approach to integration
  by parts reduction}},
\href{http://dx.doi.org/10.1016/j.physletb.2015.03.029}{{\em Phys. Lett.}
  {\bfseries B744} (2015) 101--104}
  [\href{http://arxiv.org/abs/1406.4513}{{\ttfamily arXiv:1406.4513}}]
  [\href{http://inspirehep.net/search?p=find+vonManteuffel:2014ixa}{{\ttfamily
  InSPIRE}}].

\bibitem{Peraro:2016wsq}
T.~Peraro, {\it {Scattering amplitudes over finite fields and multivariate
  functional reconstruction}},
\href{http://dx.doi.org/10.1007/JHEP12(2016)030}{{\em JHEP} {\bfseries 12}
  (2016) 030} [\href{http://arxiv.org/abs/1608.01902}{{\ttfamily
  arXiv:1608.01902}}]
  [\href{http://inspirehep.net/search?p=find+Peraro:2016wsq}{{\ttfamily
  InSPIRE}}].

\bibitem{Kosower:2018obg}
D.~A. Kosower, {\it {Direct Solution of Integration-by-Parts Systems}},
\href{http://dx.doi.org/10.1103/PhysRevD.98.025008}{{\em Phys. Rev.} {\bfseries
  D98} (2018) 025008} [\href{http://arxiv.org/abs/1804.00131}{{\ttfamily
  arXiv:1804.00131}}]
  [\href{http://inspirehep.net/search?p=find+Kosower:2018obg}{{\ttfamily
  InSPIRE}}].

\bibitem{Wang:2019mnn}
Y.~Wang, Z.~Li, and N.~Ul~Basat,
{\it {Direct Reduction of Amplitude}},
\href{http://dx.doi.org/10.1103/PhysRevD.101.076023} {{\em Phys. Rev. D}
  {\bfseries 101} (2020) 076023}
  [\href{http://arxiv.org/abs/1901.09390}{{\ttfamily arXiv:1901.09390}}]
  [\href{http://inspirehep.net/search?p=find+Wang:2019mnn}{{\ttfamily
  InSPIRE}}].

\bibitem{Mastrolia:2018uzb}
P.~Mastrolia and S.~Mizera, {\it {Feynman Integrals and Intersection Theory}},
\href{http://dx.doi.org/10.1007/JHEP02(2019)139}{{\em JHEP} {\bfseries 02}
  (2019) 139} [\href{http://arxiv.org/abs/1810.03818}{{\ttfamily
  arXiv:1810.03818}}]
  [\href{http://inspirehep.net/search?p=find+Mastrolia:2018uzb}{{\ttfamily
  InSPIRE}}].

\bibitem{Frellesvig:2019kgj}
H.~Frellesvig, F.~Gasparotto, S.~Laporta, M.~K. Mandal, P.~Mastrolia,
  L.~Mattiazzi, and S.~Mizera, {\it {Decomposition of Feynman Integrals on the
  Maximal Cut by Intersection Numbers}},
\href{http://dx.doi.org/10.1007/JHEP05(2019)153}{{\em JHEP} {\bfseries 05}
  (2019) 153} [\href{http://arxiv.org/abs/1901.11510}{{\ttfamily
  arXiv:1901.11510}}]
  [\href{http://inspirehep.net/search?p=find+Frellesvig:2019kgj}{{\ttfamily
  InSPIRE}}].

\bibitem{Frellesvig:2019uqt}
H.~Frellesvig, F.~Gasparotto, M.~K. Mandal, P.~Mastrolia, L.~Mattiazzi, and
  S.~Mizera,
{\it {Vector Space of Feynman Integrals and Multivariate Intersection
  Numbers}},
  \href{http://dx.doi.org/10.1103/PhysRevLett.123.201602} {{\em Phys. Rev. Lett}
  {\bfseries 123} (2019) 201602}
   [\href{http://arxiv.org/abs/1907.02000}{{\ttfamily
  arXiv:1907.02000}}]
  [\href{http://inspirehep.net/search?p=find+Frellesvig:2019uqt}{{\ttfamily
  InSPIRE}}].

\bibitem{Klappert:2019emp}
J.~Klappert and F.~Lange, {\it {Reconstructing Rational Functions with
  $\texttt{FireFly}$}},
\href{http://dx.doi.org/10.1016/j.cpc.2019.106951}{{\em Comput. Phys. Commun.}
  {\bfseries 247} (2020) 106951}
  [\href{http://arxiv.org/abs/1904.00009}{{\ttfamily arXiv:1904.00009}}]
  [\href{http://inspirehep.net/search?p=find+Klappert:2019emp}{{\ttfamily
  InSPIRE}}].

\bibitem{Gluza:2010ws}
J.~Gluza, K.~Kajda, and D.~A. Kosower, {\it {Towards a Basis for Planar
  Two-Loop Integrals}},
\href{http://dx.doi.org/10.1103/PhysRevD.83.045012}{{\em Phys. Rev.} {\bfseries
  D83} (2011) 045012} [\href{http://arxiv.org/abs/1009.0472}{{\ttfamily
  arXiv:1009.0472}}]
  [\href{http://inspirehep.net/search?p=find+Gluza:2010ws}{{\ttfamily
  InSPIRE}}].

\bibitem{Schabinger:2011dz}
R.~M. Schabinger, {\it {A New Algorithm For The Generation Of
  Unitarity-Compatible Integration By Parts Relations}},
\href{http://dx.doi.org/10.1007/JHEP01(2012)077}{{\em JHEP} {\bfseries 01}
  (2012) 077} [\href{http://arxiv.org/abs/1111.4220}{{\ttfamily
  arXiv:1111.4220}}]
  [\href{http://inspirehep.net/search?p=find+Schabinger:2011dz}{{\ttfamily
  InSPIRE}}].

\bibitem{Larsen:2015ped}
K.~J. Larsen and Y.~Zhang, {\it {Integration-by-parts reductions from unitarity
  cuts and algebraic geometry}},
\href{http://dx.doi.org/10.1103/PhysRevD.93.041701}{{\em Phys. Rev.} {\bfseries
  D93} (2016) 041701} [\href{http://arxiv.org/abs/1511.01071}{{\ttfamily
  arXiv:1511.01071}}]
  [\href{http://inspirehep.net/search?p=find+Larsen:2015ped}{{\ttfamily
  InSPIRE}}].

\bibitem{Boehm:2018fpv}
J.~B\"ohm, A.~Georgoudis, K.~J. Larsen, H.~Sch\"onemann, and Y.~Zhang, {\it
  {Complete integration-by-parts reductions of the non-planar hexagon-box via
  module intersections}},
\href{http://dx.doi.org/10.1007/JHEP09(2018)024}{{\em JHEP} {\bfseries 09}
  (2018) 024} [\href{http://arxiv.org/abs/1805.01873}{{\ttfamily
  arXiv:1805.01873}}]
  [\href{http://inspirehep.net/search?p=find+Boehm:2018fpv}{{\ttfamily
  InSPIRE}}].

\bibitem{Bendle:2019csk}
D.~Bendle, J.~Boehm, W.~Decker, A.~Georgoudis, F.-J. Pfreundt, M.~Rahn,
  P.~Wasser, and Y.~Zhang,
{\it {Integration-by-parts reductions of Feynman integrals using Singular and
  GPI-Space}},
\href{http://dx.doi.org/10.1007/JHEP02(2020)079} {{\em JHEP}
 {\bfseries 02} (2020) 079}
 [\href{http://arxiv.org/abs/1908.04301}{{\ttfamily
  arXiv:1908.04301}}]
  [\href{http://inspirehep.net/search?p=find+Bendle:2019csk}{{\ttfamily
  InSPIRE}}].

\bibitem{Chawdhry:2018awn}
H.~A. Chawdhry, M.~A. Lim, and A.~Mitov, {\it {Two-loop five-point massless QCD
  amplitudes within the integration-by-parts approach}},
\href{http://dx.doi.org/10.1103/PhysRevD.99.076011}{{\em Phys. Rev.} {\bfseries
  D99} (2019) 076011} [\href{http://arxiv.org/abs/1805.09182}{{\ttfamily
  arXiv:1805.09182}}]
  [\href{http://inspirehep.net/search?p=find+Chawdhry:2018awn}{{\ttfamily
  InSPIRE}}].

\bibitem{Borowka:2016ehy}
S.~Borowka, N.~Greiner, G.~Heinrich, S.~P. Jones, M.~Kerner, J.~Schlenk,
  U.~Schubert, and T.~Zirke, {\it {Higgs Boson Pair Production in Gluon Fusion
  at Next-to-Leading Order with Full Top-Quark Mass Dependence}},
  \href{http://dx.doi.org/10.1103/PhysRevLett.117.079901,
  10.1103/PhysRevLett.117.012001}{{\em Phys. Rev. Lett.} {\bfseries 117} (2016)
  012001} [\href{http://arxiv.org/abs/1604.06447}{{\ttfamily
  arXiv:1604.06447}}]
  [\href{http://inspirehep.net/search?p=find+Borowka:2016ehy}{{\ttfamily
  InSPIRE}}].
[Erratum: Phys. Rev. Lett.117,no.7,079901(2016)].

\bibitem{Jones:2018hbb}
S.~P. Jones, M.~Kerner, and G.~Luisoni, {\it {Next-to-Leading-Order QCD
  Corrections to Higgs Boson Plus Jet Production with Full Top-Quark Mass
  Dependence}},
\href{http://dx.doi.org/10.1103/PhysRevLett.120.162001}{{\em Phys. Rev. Lett.}
  {\bfseries 120} (2018) 162001}
  [\href{http://arxiv.org/abs/1802.00349}{{\ttfamily arXiv:1802.00349}}]
  [\href{http://inspirehep.net/search?p=find+Jones:2018hbb}{{\ttfamily
  InSPIRE}}].

\bibitem{Liu:2018dmc}
X.~Liu and Y.-Q. Ma,
{\it {Determine Arbitrary Feynman Integrals by Vacuum Integrals}},
\href{http://dx.doi.org/10.1103/PhysRevD.99.071501} {{\em Phys. Rev. D}
  {\bfseries 99} (2019) 071501}
  [\href{http://arxiv.org/abs/1801.10523}{{\ttfamily arXiv:1801.10523}}]
  [\href{http://inspirehep.net/search?p=find+Liu:2018dmc}{{\ttfamily
  InSPIRE}}].

\bibitem{Liu:2017jxz}
X.~Liu, Y.-Q. Ma, and C.-Y. Wang, {\it {A Systematic and Efficient Method to
  Compute Multi-loop Master Integrals}},
\href{http://dx.doi.org/10.1016/j.physletb.2018.02.026}{{\em Phys. Lett.}
  {\bfseries B779} (2018) 353--357}
  [\href{http://arxiv.org/abs/1711.09572}{{\ttfamily arXiv:1711.09572}}]
  [\href{http://inspirehep.net/search?p=find+Liu:2017jxz}{{\ttfamily
  InSPIRE}}].

\bibitem{Denner:2005nn}
Denner, Ansgar and Dittmaier, S., {\it {Reduction schemes for one-loop tensor integrals}},
\href{http://dx.doi.org/10.1016/j.nuclphysb.2005.11.007}{{\em Nucl. Phys.}
  {\bfseries B734} (2006) 62--115}
  [\href{http://arxiv.org/abs/hep-ph/0509141}{{\ttfamily hep-ph/0509141}}]
  [\href{http://inspirehep.net/search?p=find+Denner:2005nn}{{\ttfamily
  InSPIRE}}].

\bibitem{Gehrmann:1999as}
T.~Gehrmann and E.~Remiddi, {\it {Differential equations for two loop four
  point functions}},
\href{http://dx.doi.org/10.1016/S0550-3213(00)00223-6}{{\em Nucl. Phys.}
  {\bfseries B580} (2000) 485--518}
  [\href{http://arxiv.org/abs/hep-ph/9912329}{{\ttfamily hep-ph/9912329}}]
  [\href{http://inspirehep.net/search?p=find+Gehrmann:1999as}{{\ttfamily
  InSPIRE}}].

\bibitem{Lee:2008tj}
R.~N. Lee, {\it {Group structure of the integration-by-part identities and its
  application to the reduction of multiloop integrals}},
\href{http://dx.doi.org/10.1088/1126-6708/2008/07/031}{{\em JHEP} {\bfseries
  07} (2008) 031} [\href{http://arxiv.org/abs/0804.3008}{{\ttfamily
  arXiv:0804.3008}}]
  [\href{http://inspirehep.net/search?p=find+Lee:2008tj}{{\ttfamily InSPIRE}}].

\bibitem{www:reduction}
{\it
  \url{http://faculty.pku.edu.cn/yqma/en/article/40726/content/1659.htm}},
  .

\end{thebibliography}
\end{document}